\documentclass[fleqn,12pt,twoside]{article}
\usepackage{espcrc1}
\usepackage{epsf}
\usepackage{pslatex}

\usepackage{graphicx}
\usepackage[figuresright]{rotating}

\newcommand{\AmS}{{\protect\the\textfont2
  A\kern-.1667em\lower.5ex\hbox{M}\kern-.125emS}}
\newcommand{\be}{\begin{eqnarray}}
\newcommand{\ee}{\end{eqnarray}}

\newcommand{\raf}[1]{(\ref{#1})}
\newcommand{\ave}[1]{\left\langle #1 \right\rangle}
\newcommand{\ebe}{E-by-E\ }

\title{Event-by-event fluctuations and the QGP}

\author{V. Koch\address[MCSD]{Nuclear Science Division, Lawrence Berkeley
    National Laboratory, \\ 
    1 Cyclotron Road, Berkeley, CA 94720, USA},
  M. Bleicher\address[FFM]{Institut f\"ur Theoretische Physik,
    Universit\"at Frankfurt\\ 60054 Frankfurt am Main , Germany},
  and S. Jeon \address[mcgill]{Phyics Department, McGill University\\
    Montreal, QC H3A-2T8
    Canada}  
  }
\begin{document}

\maketitle

\begin{abstract}
We discuss the physics underlying event-by-event fluctuations in
  relativistic heavy ion collisions. We will emphasize how the fluctuations
  of particle ratios can be utilized to explore the properties of the matter 
  created in these
  collisions. In particular, we will argue that the fluctutions of the ratio of
  positively over negatively charged particles may serve as a unique
  signature for the Quark Gluon Plasma.
\end{abstract}

\section{INTRODUCTION}
Any physical quantity measured in experiment is subject to fluctuations.
In general, these fluctuations depend on the properties of the system
under study (in the case at hand, on the properties of a fireball created 
in a heavy ion collision) and may contain important information about the 
system.

The original motivation for event-by-event (\ebe) studies in ultra relativistic
heavy ion collisions has been to find indications for 
distinct event classes. In
particular it was hoped that one would find events which would carry the
signature of the Quark Gluon Plasma. First pioneering experiments in this
direction have been carried out by the NA49 collaboration 
\cite{NA49}. They analysed the \ebe fluctuations of the mean transverse
momentum as well as the kaon to pion ratio (see Fig. \raf{fig1}). 
Both observables, however, did not
show any indication for two or more distinct event classes. Moreover, the
observed fluctuations in both cases were consistent with pure statistical
fluctuations. 

On the theoretical side, the subject of \ebe fluctuations has recently gained
considerable interest. Several methods to distinguish between statistical and
dynamic fluctuations have been devised \cite{mrow,voloshin}.
Furthermore the influence of hadronic resonances and
possible phase transitions has been investigated
\cite{shuryak,jeon1,BH99,heiselberg_jackson,heiselberg_report}. All these
theoretical considerations assume that the observed fluctuations will be
Gaussian and thus the physics information will be in the width of the
Gaussian. In \cite{bialas} one of us has shown, that in this case, the
information contained in \ebe fluctuations is that of a two particle density,
which can alternatively be measured using a two arm spectrometer. This
observation implies that possibly interesting fluctuations can be observed and
verified by a whole array of detectors not only large acceptance ones.

\begin{figure}[htb]
\epsfysize=6cm
\centerline{\epsffile{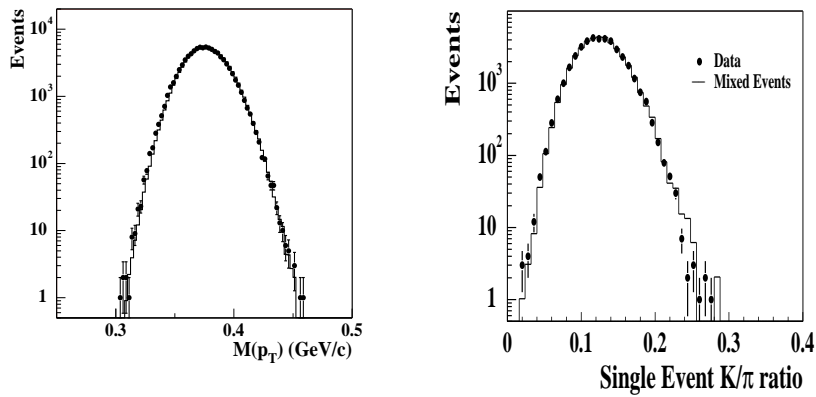}}
\caption[]{
Results from the NA49 collaboration for the \ebe fluctuations of the mean
transverse momentum as well as the kaon to pion ratio \cite{NA49}
}
\label{fig1}
\end{figure}

\section{Event-by-event fluctuations} 
Fluctuations have contributions of different nature.
First there are the `trivial' statistical fluctuations due to a finite 
number of particles used to define a particular observable in a given event.  
Second there are dynamical fluctuations. Those carry the information about the
properties of the system. In the special case of heavy ion collisions, there
is another source of `trivial' fluctuations, namely the fluctuations of the
volume. All these fluctuations contribute to \ebe fluctuations. The obvious
challenge, therefore, is to separate the `trivial' from the interesting
components.

\subsection{Fluctuations in thermal systems}
In a thermal system, the fluctuations of a quantity are proportional to its
susceptibility, which is the second derivative of the appropriate free energy
with respect to the conjugate variable. For example the fluctuations of the
charge are given by
\be
\ave{\delta Q} = - T \frac{\partial^2 F}{\partial \mu_Q^2} = V T \chi_Q
\label{charge_suscept}
\ee
where $\mu_Q$ is the charge chemical potential, $T$ the temperature and $V$
the volume of the system. $\chi_Q$ is the charge susceptibility. 
Here we will mostly concentrate on the charge fluctuations. Therefore, let us
discuss the charge susceptibility in some more detail. First, as shown in
\cite{Kapusta}, the charge susceptibility is directly related to the electric
screening mass in a system. Furthermore, it is given by the zero momentum limit
of the static current-current correlation function \cite{Kapusta}
\be
\chi_Q = \Pi_{00}(\omega=0, k\rightarrow 0)
\ee

which can be, 
and has been, evaluated in Lattice QCD \cite{Lattice,gupta} as well as in
effective hadronic models \cite{Kapusta,Song,Raju}.

\subsection{Fluctuations in Heavy Ion Collisions}
The systems created in a heavy ion collisions carry another important, but
rather uninteresting source of fluctuations namely the fluctuations of the
volume. Even for the tightest centrality selection the impact parameter and
thus the volume of the system  still fluctuates considerably. 
These volume fluctuations may even dominate.
For example let us consider the fluctuation of the number of particles of a
certain species. In a given event the number of particles is given by
\be
N = \rho V
\ee
where $\rho$ is the particle density and $V$ the volume of the system created
in this event. Thus the fluctuations have contributions from both 
the density fluctuations {\em and} the fluctuations of the volume
\be
\ave{(\delta N)^2} = \ave{(\delta \rho)^2} \ave{V^2} + 
                       \ave{\rho^2} \ave{(\delta V)^2}.
\ee
The interesting physics of course resides in the fluctuations of the
density. Consequently, one needs to consider so called `intensive' quantities,
which do no explicitely depend on the volume. One such observable is the ratio
of particle abundances, and this is what we will concentrate on here.

\subsection{Fluctuations of particle ratios}
\label{rat_fluct}
As discussed in the previous section, the fluctuations of particle ratios
should be independent of volume fluctuations. This is certainly true if one
looks at similar particles such as $\pi^+$ and $\pi^-$, where the freeze out
volumes are expected to be the same. Some residual volume fluctuations may be
present if one considers ratios of particles with different quantum numbers
such as the $K/\pi$ ratio, but they still should be small.
Let us define the particle ratio $R_{12}$ of two particle species $N_1$ and $N_2$
\be
R_{12} = \frac{N_1}{N_2}
\ee
The fluctuations of this ratio are then given by \cite{mrow,jeon1,BH99}
\be
\frac{(\delta R_{12})^2}{\ave{R_{12}}^2}  
& = &
\left(\frac{\ave{(\delta N_{1})^2}}{\ave{N_{1}}^{2}} + 
\frac{\ave{(\delta N_{1})^2}}{\ave{N_{2}}^{2}} 
\right. 
-2
\left. 
\frac{\ave{\delta N_{1} \delta N_{2}}}{\ave{N_{1}}\ave{N_{2}}} \right) .
\label{eq:ratio-fluct}
\ee 
The last
term in Eq.~(\ref{eq:ratio-fluct}) takes into account correlations between the
particles of type 1 and type 2. This term will be important if both particle
types originate from the decay of one and the same resonance. For example, in
case of the $\pi^+/\pi^-$ratio, the $\rho_0$, $\omega$ etc.  
contribute to these correlations. 
Also this term is responsible to cancel out all volume fluctuations
\cite{jeon1}.

Let us note that the effect of the
correlations introduced by the resonances should be most visible
when \( \ave{N_{1}}\simeq \ave{N_{2}} \).
On the other hand, 
when \( \ave{N_{2}}\, \gg \, \ave{N_{1}} \),
as in the $K$ to $\pi$ ratio, 
the fluctuation is dominated by the less abundant particle 
type and the resonances feeding into it. The correlations are then very hard
to extract. In \cite{jeon1} it was shown that in case of the $K/\pi$-ratio 
resonances and quantum statistics
give rise to deviations from the statistical value of at most 2~\%, in
agreement with experiment \cite{NA49}.

As pointed out in \cite{jeon1} the measurement of particle ratio fluctuations
can provide important information about the abundance of resonances at chemical
freeze out, and thus provides a crucial test for the picture emerging from the
systematics of single particle yields \cite{stachel}. In particular the
fluctuations of the $\pi^+/\pi^-$-ratio should be reduced by about 30 \% as
compared to pure statistics due to the presence of hadronic resonances with
decay channels into a $\pi^+$-$\pi^-$-pair at chemical freeze out.  About 50~\%
of the correlations originate from the decay of the $\rho_0$ and the $\omega$
mesons. Thus the fluctuations provide a complementary measurement to the
dileptons. 

\section{Charge fluctuations}
 Measuring the charge fluctuations or more precisely the charge fluctuations
 per unit degree of freedom of the system created in a heavy ion collision
 would tell us immediately if we have created a system of quarks and gluons
 \cite{jeon2} (see also \cite{mueller}).
 The point is that in the QGP phase, the unit of charge is $1/3$
 while in the hadronic phase, the unit of charge is 1.
 The net charge, of course does not depend on such subtleties, but the
 fluctuation in the net charge depends on the {\em squares} of the 
 charges and hence are strongly dependent on which phase it originates
 from.  However, as discussed in the previous section, 
 measuring the charge fluctuation itself is plagued by
 systematic uncertainties such as volume fluctuations, which can be avoided if
 one considers ratio fluctuations. 
 The task is then to find a suitable ratio whose fluctuation 
 is easy to measure and simply related to the net charge fluctuation. 
 
 The obvious candidate is the ratio $F = Q/N_{\rm ch}$ 
 where $ Q = N_+ - N_-$  is the net charge and  $ N_{\rm ch} = N_+ + N_-$
 is the total charge  multiplicity. 
 Instead of using $F$, however, it is simpler to consider the 
 the charge ratio $R = N_+/N_-$. 
 If $\ave{N_{\rm ch}} \gg \ave{Q}$, i.e.  $\left| F \right| \ll 1$ the
 fluctuations of $F$ are related to those of $R$ by
 \be
 \ave{\delta R^2}
 \approx  
 4\ave{\delta F^2}
 \ee
 Furthermore in the limit of $\ave{Q}<< \ave{N_{\rm ch}}$ the fluctuations are
 dominated by the small term and we find \cite{jeon2}
 \be
 \ave{\delta F^2} \approx {\ave{\delta Q^2}\over \ave{N_{\rm ch}}^2}
 \ee
 Usually, the number of charged particles is directly related to the entropy
 generated in these collision. Thus the  observable
 \be 
 D \equiv \ave{N_{\rm ch}}\ave{\delta R^2} = 4\ave{N_{\rm ch}}\ave{\delta F^2} 
 =4 {\ave{\delta Q^2} \over \ave{N_{\rm ch}}}
 \ee
 provides a measure of the charge fluctuations per unit entropy.

 In order to see how this observable differs between a hadronic system and a
 QGP let us compare the value for $D$ in a pion gas and in a simple model
 of free quarks and gluons. 
 
 In a pion gas, the fundamental degrees of freedom are of course pions.
 Hence, $N_{\rm ch} = N_{\pi^+} + N_{\pi^-}$ and 
 using thermal distributions and disregarding correlations the charge
 fluctuations are given by 
 \be
 \ave{\delta Q^2} = \ave{\delta N_+^2} + \ave{\delta N_-^2}
 =
 w_\pi\ave{N_{\rm ch}}
 \label{eq:numerator}
 \ee
 where $ w_\pi \equiv \ave{\delta N_\pi^2}/\ave{N_\pi} $
 is slightly bigger than 1 \cite{jeon1,BH99}.
 Hence for a pion gas, 
 \be
 D_{\pi-gas} \approx 4
 \;.
 \label{eq:pion_gas}
 \ee
 In the presence of resonances, this value gets reduced by about 30 \% due to 
 the
 correlations introduced by the resonances, as discussed in section 
 \raf{rat_fluct}. 

 For a thermal system of free quarks and gluons we have in the absence
 correlation  
 \be
 \ave{\delta Q^2} = Q_u^2 w_u \ave{N_u} + Q_d^2 w_d \ave{N_d}
 \ee
 where $Q_{q}$ is the charges of the quarks 
 and $N_{q}$ denotes the number of quarks {\em and} anti-quarks.
 The constant $ w_q \equiv \ave{\delta N_q^2}/\ave{N_q}$
 is slightly smaller than 1 due to the fermionic nature of quarks.

 Relating the final charged particle multiplicity $N_{\rm ch}$ to the
 number of primordial quarks and gluons is not as simple. Using entropy
 conservation one finds \cite{jeon2}
  \be
 N_{\rm ch} \simeq {2\over 3}\left(N_g + 1.2 N_u + 1.2 N_d \right)
 \ee
 leading to 
 \be
 D_{QGP} \simeq 3/4
 \ee
 
 Actually the charge fluctuations $\ave{(\delta Q)^2}$ have been evaluated in
 lattice QCD along with the entropy density \cite{Lattice}. Using these
 results one finds
 \be
 D_{Lattice-QCD} \simeq 1 - 1.5
 \ee
 where the uncertainty results from the uncertainty of relating the entropy
 to the number of charged particles in the final state. Actually the most 
 recent  lattice result \cite{gupta} for the charge fluctuations, 
 which was obtained in the quenched approximation, is somewhat lower then 
 the result of \cite{Lattice}.

 But even using the larger value of $D = 1.5$ for the Quark Gluon Plasma,
 there is still a factor of 2 difference between a hadronic gas and the QGP,
 which should be measurable in experiment.

 \section{Observational issues and model calculations}
 The key question of course is, how can these reduced fluctuation be observed
 in the final state which consists of hadrons. Should one not expect that
 the fluctuations will be those of the hadron gas? 
 The reason, why it should be possible to see the charge fluctuations of the
 initial QGP has to do with the fact that charge is a conserved quantity. 
 Imagine one measures in each event the net charge in a given rapidity
 interval $\Delta y$ such that 
 \be
 1 \ll \Delta y \ll \Delta y_{max}
 \ee
 where $\Delta y_{max}$ is the width of the total charge distribution.
 If, as it is expected, strong longitudinal flow develops already in the
 QGP-phase, the number of charges inside the rapidity window $\Delta y$ 
 for a given event is essentially frozen in.
 And if $\Delta y \gg 1$ neither hadronization nor the subsequent collisions
 in the hadronic phase will be very effective to transport charges in and out
 of this rapidity window. Thus, the \ebe charge-fluctuations measured at the 
 end  reflect those of the initial state, when the longitudinal flow is
 developed. Ref. \cite{stephanov} arrives at the same conclusion  
 on the basis of a Fokker-Planck type equation describing the relaxation of
 the charge fluctuation in a thermal environment. In this approach the
 relaxation time for the charge fluctuations is determined by the average
 rapidity shift of the charges in a collision and by the number of
 collisions in the hadronic phase. These values can be easily extracted from a
 transport model and we find that for $\Delta y > 3$ the signal should survive
 the hadronic phase.  
  The smearing of the signal due to
 hadronization is more difficult to estimate, however. But it is rather
 unlikely that hadronization, which is a soft process, should give rise to
 large rapidity shifts. 

 Thus, in the limit of very large $\Delta y$ our proposed  signal should be
 robust. However, one should not make $\Delta y$ too large, as eventually
 charge conservation becomes relevant. Actually in the limit that 
 $ \Delta y \simeq \Delta y_{max}$ charge conservation requires that the
 fluctuations vanish. For moderate values  of $\Delta y$, one can very
 successfully correct for the effect of charge conservation, and this needs to
 be done for a meaningful interpretation of any experimental data. 
 This correction factor is given by \cite{bleicher}
 \be
 C_y = 1-
 \frac{\langle N_{ch}\rangle_{\Delta y}}{\langle N_{ch}\rangle_{\rm total}}
 \ee
Also in
case of a finite net charge, which may be due to non negligible baryon
 stopping, some additional, though small, correction needs to
be applied \cite{bleicher}.  
\be
C_{\mu} 
= \frac{\langle N_{\Delta y}^+\rangle^2}{ \langle N_{\Delta y}^-\rangle ^2}
\ee
Thus the actual observable turns out to be
\be
\tilde{D} = \frac{D}{C_y C_\mu}
\ee
In Fig. \raf{fig:bleicher} the importance of these corrections, in
 particular the charge conservation correction $C_y$ is demonstrated. There we
 compare the uncorrected observable $D$ with the corrected observable
 $\tilde{D}$ as a function of the width of the rapidity window based on a
 URQMD \cite{URQMD} simulation. The effect of
 the charge conservation is clearly visible. With increasing rapidity window
 the charge fluctuation are suppressed. Once the charge-conservation
 corrections are applied, the value for $\tilde{D}$ remains constant at 
 the value for a hadron gas of $\tilde{D} \simeq 3$. This is to be expected
 for the URQMD model, which is of hadronic nature and does not have 
 quark- gluon degrees of freedom. 

\begin{figure}[htb]
\begin{center}
\epsfysize=6cm
\epsffile{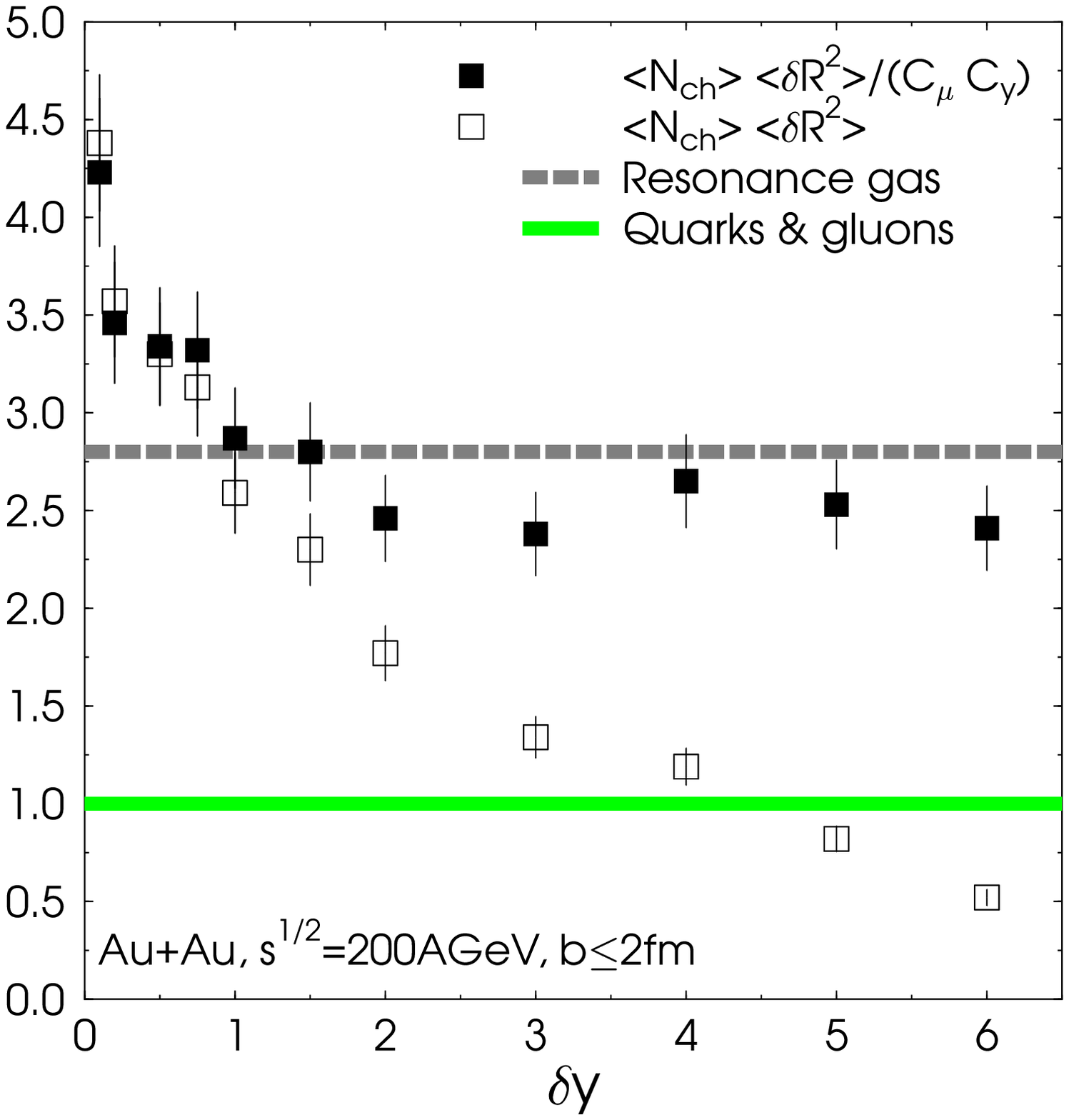}
\epsfysize=6cm
\epsffile{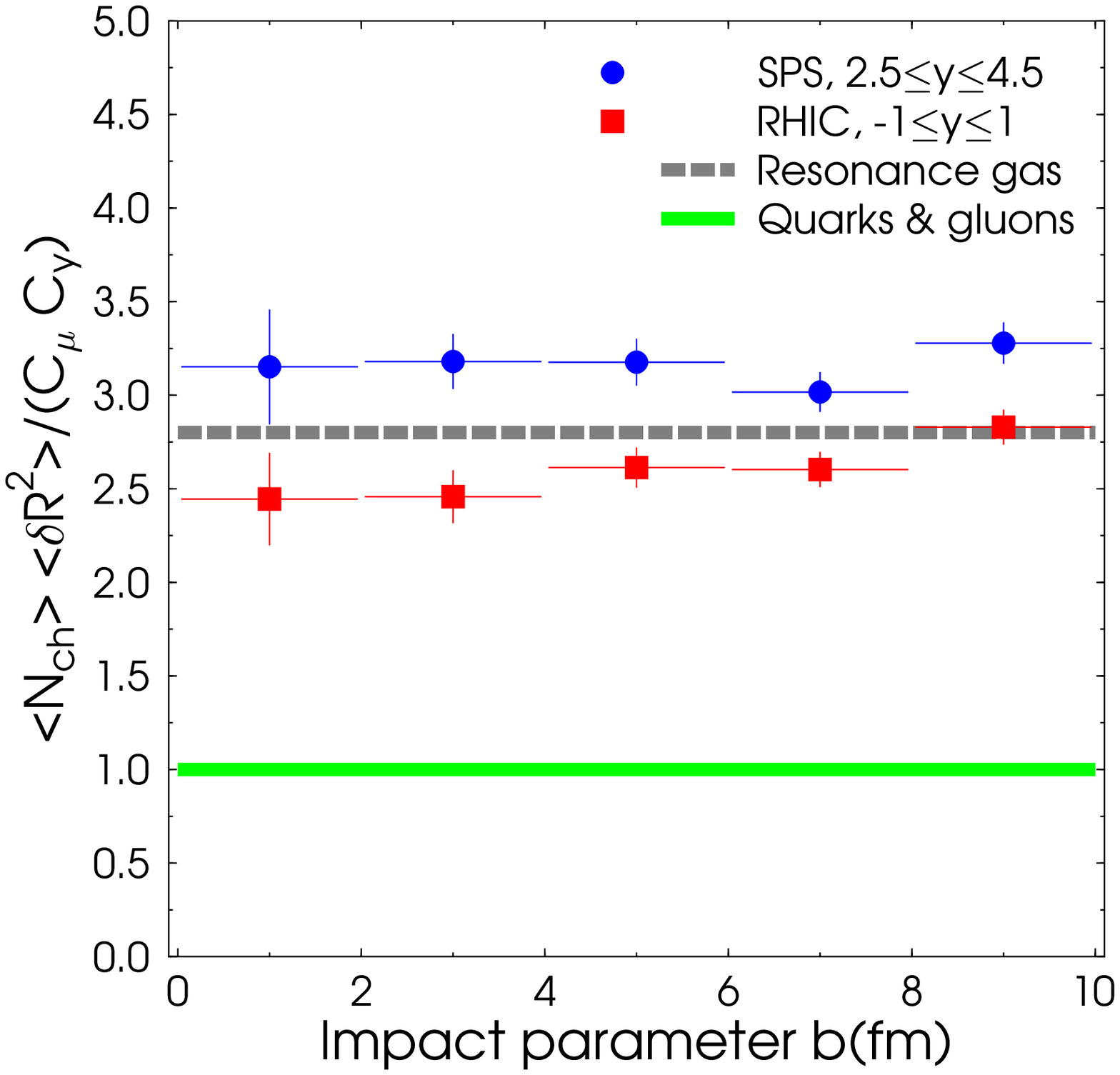}
\end{center}
\vspace*{-1cm}
\caption[]{
Charge fluctuations as a function of the size of the rapidity window(left) and
centrality (right). 
}
\label{fig:bleicher}
\end{figure}

Also for small $\Delta y$ the value is  $\tilde{D} \simeq 4$
before it drops to $\tilde{D} \simeq 3$ for $\Delta y > 1.5$. This
effect, which was already predicted in \cite{jeon1}, is simply due to the fact
that the correlation introduced by the resonances gets lost if the
acceptance window becomes too small.

In Fig. \raf{fig:bleicher}, we also 
show the predictions of the URQMD model as a
function of impact parameter for SPS as well as RHIC energies
\cite{bleicher}. 
For both
energies hardly any centrality dependence is visible and the results agree
within errors with the prediction for the hadron gas. Thus a
measurement of $\tilde{D} \simeq 1$ would clearly indicate the existence of a
QGP in the system created in these collisions.

\section{Some additional remarks}
A similar argument as for the charge fluctuations also holds for the
fluctuations of the baryon number, since in the QGP the quarks carry
fractional baryon number \cite{mueller}. The observation of reduced baryon
number fluctuations, however, is much more difficult since one has to measure
neutrons on an event by event basis. Also, so far, no obvious
{\em intensive} observable has been proposed which would be sensitive to
baryon number fluctuations.  

Certainly, a measurement of $D$ in proton-proton collisions will be necessary
in order to rule out that possible small fluctuations are simply 
due to the nature of the proton wavefunction.

In order to be able to see the reduced charge fluctuations, they have to be
imprinted into the hadronic phase. How can this be possible? As we have shown
above, the charge fluctuations of a hadronic gas are three times larger than
that of a QGP. The only way to reduce the fluctuations in a hadron gas is to
introduce additional correlations (see eq. \raf{eq:ratio-fluct}). As we have
argued above, neutral hadronic resonances are the obvious candidates, which
would lead us to predict an enhancement of neutral resonances. However, if,
in addition, we require that on average iso-spin should be conserved,
increasing the number of neutral rho mesons would also imply an increase of
the charged rho mesons and thus large charge fluctuations. This leaves us with
the iso-scalar resonances, such as eta, omega, phi etc. We thus are lead to
predict an enhancement of eta, omega and phi production, which can be readily
observed via electromagnetic decay channels. An actual calculation
based on this conjecture gives enhancement factors of the order of 5.

\section{Conclusions}
We have discussed event-by-event fluctuations  in heavy ion
collisions. These fluctuations may provide useful information  
about the properties of the matter created in these collisions, as long as the
`trivial' volume fluctuations, inherent to heavy ion collisions, can be
removed. We have argued that the fluctuations of particle ratios is not
affected by volume fluctuations.  

In particular the fluctuations of the ratio of positively over negatively
charged particles measures the charge fluctuation per degree of freedom. Due
to the fractional charge of the quarks, these are smaller in a QGP than in a
hadronic system. 

A measurement of our observable $\tilde{D} \simeq 1$ 
would provide strong evidence for the existence of a QGP in these
collisions. A measurement of $\tilde{D} \simeq 3$ on the other hand does not
rule out the creation of a QGP. There are a number of caveats (see
\cite{jeon2}), which could destroy the signal, such as unexpected large
rapidity shifts during hadronization.

\section*{Acknowledgments}
This work was supported by the Director, Office of Energy Research,
Office of High Energy and Nuclear Physics, Division of Nuclear Physics,
the Office of Basic Energy
Science, Division of Nuclear Science, of the U.S. Department of Energy under
Contract No. DE-AC03-76SF00098. 
M.B. was also supported by the Feydor Lynen
Program of the the Alexander von Humboldt Foundation.

\end{document}